%% file: globecom25_ieeeformat.tex
\documentclass[conference]{IEEEtran}
\IEEEoverridecommandlockouts
\usepackage{cite}
\usepackage{amsmath,amssymb,amsfonts}

\usepackage{algorithm}
\usepackage{algcompatible}
\usepackage{graphicx}
\usepackage{textcomp}
\usepackage{xcolor}
\usepackage{latexsym}
\usepackage{setspace}
\usepackage{bm}
\usepackage{lipsum}
\usepackage{bbm}
\usepackage{tikz}

\usepackage{amsmath,graphicx,xcolor,mathcomSTEv4}
\usepackage{mathtools}

\input{preamble}
\usepackage{accents}
\usepackage[bindingoffset=0in,left=0.65in,right=0.65in,top=0.75in,bottom=1in]{geometry}
\begin{document}

\title{Wireless Decentralized Federated Learning via Device Clustering and Inter-Cluster Link Enhancement 
\vspace{-0.3cm}
}

\author{William Weijia Zheng\textsuperscript{$\dagger$}, Hang~Liu\textsuperscript{*}, and~Ying-Jun~Angela~Zhang\textsuperscript{$\dagger$}\\
	\textsuperscript{$\dagger$}Department of Information Engineering, The Chinese University of Hong Kong, Hong Kong SAR\\
    		\textsuperscript{*}State Key Laboratory of Internet of Things for Smart City and Department of \\ Electrical and Computer Engineering, University of Macau, Macao S.A.R. \\
	Email: wjzheng@link.cuhk.edu.hk, hangliu@um.edu.mo, yjzhang@ie.cuhk.edu.hk%
\vspace{-0.35cm}


}

\maketitle

\begin{abstract}
Decentralized federated learning (DFL) dispenses with the central server of classical FL by utilizing peer-to-peer model exchanges among edge devices.  This server-free architecture enables ad-hoc, flexible distributed learning in large device-to-device (D2D) networks. However, wireless DFL converges slowly because peer-to-peer model aggregation incurs high delays and errors. Each DFL training round involves many-to-many gradient sharing over wireless channels, resulting in uncoordinated channel access, large communication errors from stragglers, and slow model consensus, especially in large-scale D2D networks with pronounced clustering structures.
We address these aggregation bottlenecks by provisioning a few \emph{reliable backhaul links} at straggling nodes to enhance network connectivity. Building on this idea, our \emph{budget-aware, cluster-centric} DFL framework first partitions the network into densely connected clusters, and then allocates the limited backhaul budget to selected cluster heads. The resulting two-tier protocol executes fast, parallel model aggregation within clusters and infrequent inter-cluster exchanges among the heads, yielding an $\mathcal{O}(1/t)$ convergence rate in $t$ iterations. Numerical experiments on image-classification tasks confirm that our approach accelerates convergence compared to state-of-the-art DFL baselines with only a few strategically placed backhaul links.
\end{abstract}


\section{Introduction}
\label{sec:intro}

Federated learning (FL) has emerged as a compelling framework for distributed training of artificial-intelligence (AI) models at the network edge, where both data and computation are distributed over edge devices \cite{fedavg17}.  Classical FL relies on a central server (e.g., a base station or edge server) to coordinate model aggregation: edge devices train local models and periodically upload their models or gradients to the server for global consensus.  Although effective, this server-centric architecture becomes impractical whenever device-to-server connections are unavailable, unreliable, or undesirable, particularly in ad hoc networks \cite{FL_at_scale}.  \emph{Decentralized FL} (DFL) overcomes this limitation by replacing the server with peer-to-peer gossip exchanges \cite{XB03}, thereby aligning with the connectivity patterns and privacy requirements of large-scale device-to-device (D2D) networks and the Internet-of-Things (IoT) \cite{FL_safe1}.

A growing body of work confirms that model communication and aggregation, rather than local model computation, dominate the runtime of FL systems \cite{9084352}.  Exchanging high-dimensional model parameters across rate-limited links is extremely costly for large networks. The challenge is amplified in DFL as every device must alternately transmit to and receive from its neighbors, turning the ``many-to-one" uplink of classical FL into \emph{``many-to-many"} communication \cite{chocoSGD}. The absence of a coordinating server further complicates decentralized multiple-access control, leading to collisions, long delays, and ultimately slower convergence.

To mitigate the model aggregation bottlenecks, recent research advocates \emph{over-the-air} (OTA) computation as a scalable solution. By exploiting the signal-superposition property of the wireless multiple-access channel, OTA computation sums gradients ``in the air" by transmit scaling, allowing all devices to transmit simultaneously to a single receiver over the same channel \cite{FL_ota_2020}.  When extended to DFL \cite{shi2021ota-dfl,DSGD_OTA20,XSB21}, OTA aggregation reduces the peer-to-peer gossip delay among $N$ devices from $\mathcal{O}(N^{2})$ transmissions to $\mathcal{O}(N)$.

Despite its promise, OTA-based DFL still faces three key hurdles: (\emph{a}) without a server, devices struggle to schedule concurrent transmissions and receptions \cite{shi2021ota-dfl}; (\emph{b}) OTA aggregation relies on transmitter-side channel inversion to align the received gradients, forcing all transmitters to scale to the stragglers with weakest links, which inflates noise and slows convergence \cite{HL21}; and (\emph{c}) large D2D networks typically exhibit “small-world’’ structure, where dense clusters bridged by only a few inter-cluster edges. As a consequence, the sparsely connected clusters throttle global consensus \cite{chocoSGD}.

These observations suggest harnessing the natural clustering of ``small-world" networks to localize OTA aggregation within dense clusters and reinforce the few critical long-range edges. This insight motivates supplementing DFL with a few \emph{extra reliable backhaul links} to assist the sparsely connected stragglers, bridge otherwise isolated clusters, and, in turn, accelerate model aggregation. Specifically, we propose a \emph{budget-aware, cluster-centric} DFL framework that (\emph{a}) partitions the network into densely connected clusters, (\emph{b}) identifies straggling devices whose poor connections dominate aggregation error, and (\emph{c}) deploys a small set of extra reliable backhaul links between selected cluster heads.  The resulting two-tier protocol performs fast, head-centric OTA aggregation inside each cluster and low-frequency inter-cluster exchanges among the heads, achieving an $\mathcal{O}(1/t)$ convergence rate in the optimality-gap bound after $t$ iterations.  Finally, we develop a low-complexity clustering algorithm that minimizes the convergence rate bound under link-deployment constraints using only long-term channel statistics. Notably, our approach generalizes the spirit of hierarchical FL to a fully decentralized setting, removing the need for dedicated cloud  servers when partitioning users. Moreover, it naturally adapts to heterogeneous ad-hoc networks: clusters can leverage existing infrastructure, such as routers or factory controllers, as pre-established inter-cluster links, further trimming costs.



\section{System Model and Preliminaries}
\label{sec:system_model}


In this work, we consider a DFL system operating over a D2D network, where $N$ \emph{single-antenna} IoT devices collaboratively train a shared AI model using their local datasets. We assume that these devices form a connected network through wireless connections where there is a routing path connecting any two devices. The global objective is to minimize 
\begin{equation}
    \min_{\xv \in \mathbb{R}^d} f(\xv)=\min_{\xv \in \mathbb{R}^d} \sum_{n=1}^N \frac{|\mathcal{D}_n|}{D} f_n(\xv),
    \label{eq: grand objective}
\end{equation}
where $\xv \in \mathbb{R}^d$ represents the model parameters of dimension $d$, $\mathcal{D}_n$ is the dataset on device $n$, $D=\sum_n |\mathcal{D}_n|$ is the total number of data, and $f_n(\cdot)$ is the local loss of device $n$. Specifically, the local loss is evaluated over all samples in the local dataset $\mathcal{D}_n$ as
\begin{equation}
    f_n(\xv)=\frac{1}{|\mathcal{D}_n|} \sum_{i=1}^{|\mathcal{D}_n|} f_n(\xv;\xi_{n,i}),
    \label{eq: local objective}
\end{equation}
where $\xi_{n,i}$ denotes the $i$-th sample in $\mathcal{D}_n$.

DFL proceeds iteratively with on-device local training and inter-device model aggregation. 
Due to the network connectivity and data privacy concerns, we consider model communication occurs only among neighboring devices over wireless fading channels, with no central server involved. This \emph{fully decentralized} architecture aligns naturally with dense D2D IoT systems, where devices primarily share local connections rather than relying on a base station.

\subsection{Preliminaries on Conventional DFL Approaches}

Conventional DFL algorithms, e.g., \cite{chocoSGD,XSB21}, typically employ gossip algorithms to disseminate and aggregate local information to enforce global model consensus. For each training round $t=1,2,\ldots,T$, a standard gossip-based DFL procedure comprises the following steps:
\begin{itemize}[leftmargin=0.3cm]
    \item \textbf{Local gradient computation:} Given its current model $\xv_n^{(t)}$, device $n$ draws a mini\mbox{-}batch $\mathcal{B}_{n}^{(t)}\subset\mathcal{D}_n$ and runs mini\mbox{-}batch stochastic gradient descent (SGD) to obtain the stochastic gradient vector ${\nabla} f_{n}(\xv_n^{(t)};\mathcal{B}_{n}^{(t)})$.

    \item \textbf{Gradient exchange:} Let $\mathcal{N}_n$ denote the neighbors of device $n$ that share direct wireless links with it. Device $n$ transmits its model update with respect to $\hat{\xv}_n^{(t)}$ to all $m\in\mathcal{N}_n$ and simultaneously receives their updates. Due to communication impairments (e.g., wireless fading, compression loss), the gradient received from $m$ arrives perturbed. We denote the received model update from $m$ to $n$ by $\gv_{m\to n}^{(t)}$.

    \item \textbf{Model aggregation:} After collecting $\{\gv_{m\to n}^{(t)}\}_{m\in\mathcal{N}_n}$ from its neighbors, each device $n$ updates its model via (cf. \cite[Alg. 2, Line 9]{chocoSGD})
    \begin{align}
        &\xv_n^{(t+1)}= \xv_n^{(t+\frac{1}{2})}
        + \gamma \sum_{m\in\mathcal{N}_n \cup \{n\}}w_{nm}  (\widehat\xv_m^{(t+1)}- \widehat\xv_n^{(t+1)}),\label{Eq3}
    \end{align}
    where  $w_{nn}\geq 0$ and $w_{nm}\geq 0$ are the mixing weights controlling the balance between local and neighboring model information and satisfying $w_{nn}+\sum_{m\in \mathcal{N}_n} w_{nm}=1$, $\gamma$ denotes the consensus stepsize, 
    $\xv_n^{(t+\frac{1}{2})}=\xv_n^{(t)}-\eta_t {\nabla} f_{n}(\xv_n^{(t)};\mathcal{B}_{n}^{(t)})$ is the locally updated model with stepsize $\eta_t$, and $\widehat \xv_m^{(t)}=\xv_m^{(0)}-\sum_{t^\prime=1}^{t-1} \gv_{m\to n}^{(t^\prime)}$ is the running estimate of $m$’s model reconstructed at device $n$ from the received model updates.
\end{itemize}
It has been extensively documented that exchanging gradient information constitutes the \emph{critical bottleneck} in DFL systems. Iterative transmission and reception of local gradients over every pair of neighboring devices impose a prohibitively large latency and introduce communication errors that slow down model consensus and hinder convergence. Our work aims to address this bottleneck through an \emph{efficient device scheduling and gradient aggregation algorithm} tailored to fully decentralized networks.

\subsection{Communication Model}
OTA computation is a scalable ``many-to-one" gradient uploading scheme for dense networks. Since our proposed method employs OTA schemes to facilitate inter-device aggregation, we summarize the underlying channel model and OTA aggregation methods as follows.

For every existing wireless link between devices $m$ and $n$, the instantaneous channel coefficient at iteration $t$ is assumed to follow a Rayleigh fading model, as
\begin{align}
    h_{mn}^{(t)}=\sqrt{\alpha_{mn}}\,\beta_{mn}^{(t)}, \qquad n\in[N],\; m\in\mathcal{N}_n,\label{eq_channel}
\end{align}
where $\alpha_{mn}$ is the large-scale fading (assumed constant during training) and $\beta_{mn}^{(t)}\sim \mathcal{CN}(0,1)$ is the small-scale fading. Because large-scale coefficients vary slowly, $\{\alpha_{mn}\}$ are treated as known \emph{a priori} to the DFL system design. In contrast, the fast-varying coefficients $\{\beta_{mn}^{(t)}\}$ are estimated at the start of each training iteration. Therefore, instantaneous channel state information (CSI) is available for the OTA transceiver design but not for DFL device scheduling.

Consider a fixed iteration $t$, where the index $t$ is omitted when clear. Device $n$ aims to receive and aggregate the stochastic gradients $\{{\nabla} f_{m}(\xv_m;\mathcal{B}_{m})\}_{m\in\mathcal{N}_n}$ from its neighbors. To achieve this, each sender normalizes its gradient by the empirical mean and standard deviation, scales the normalized vector with a transmit scalar, and transmits the vectors entry-by-entry over the shared channel. The transmitted signal from device $m\in\mathcal{N}_n$ to $n$ is given by
\begin{align}
    \zv_{m\to n}=q_{m\to n}\,
    \frac{{\nabla} f_{m}(\xv_m;\mathcal{B}_{m})-\texttt{mean}\cdot\boldsymbol{1}}{\texttt{std}},\label{eq5}
\end{align}
where $q_{m\to n}\in\mathbb{C}$ is a complex scaling factor and \texttt{mean}, \texttt{std} are the common empirical statistics of all the gradients in this iteration.\footnote{Following \cite{HL21}, these two scalar statistics are computed, shared, and aggregated among all the devices with negligible overhead at the beginning of each aggregation round.} The average per-entry power is constrained by $\Ebb[\lVert \zv_{m\to n}\rVert^2/d]=|q_{m\to n}|^{2}\le P_{0}$.

Device $n$ receives a noisy superposition $\yv_{n}=\sum_{m\in\mathcal{N}_n}h_{mn}\zv_{m\to n}+\nv_{n}$, where $\nv_{n}$ is additive white Gaussian noise. By applying a linear receive scalar $\omega_{n}$, the device estimates
\begin{align}
       \sum_{m\in\mathcal{N}_n}w_{nm} \gv^{(t)}_{m\to n} \approx  \frac{\texttt{std}}{\omega_{n}}\yv_{n}+\texttt{mean}\cdot\boldsymbol{1},\label{eq_otareceiver}
\end{align}
which is then used in the local update step in \eqref{Eq3}.

Following \cite{FL_ota_2020, HL21}, the transceiver scalars that minimize the mean-squared error (MSE) of the aggregated gradients under the maximum power $P_0$ are given by
\begin{equation}
    \omega_{n}=\sqrt{P_{0}}\min_{m\in\mathcal{N}_n}\frac{|h_{mn}|}{w_{nm}},
    q_{m\to n}=\omega_{n}w_{nm}\frac{h_{mn}^{*}}{|h_{mn}|^{2}},\label{eq6}
\end{equation}
where $h_{mn}^{*}$ denotes the complex conjugate of $h_{mn}$.

\subsection{Challenges in DFL Model Aggregation}

Applying OTA aggregation to fully decentralized settings introduces several obstacles. In DFL, each device must both \emph{transmit to} and \emph{receive from} its neighbors, significantly complicating the gradient exchange procedure. Concretely, OTA-based DFL faces three key challenges, which hamper consensus and slow convergence:

\begin{itemize}[leftmargin=0.3cm]
    \item \textbf{Multiple-access control:} Classic OTA aggregation assumes a single server that orchestrates channel access. In DFL, no such coordinator exists, making it difficult to schedule the channel access, so that every device can fully transmit to, and receive from, all of its neighbors without collisions.
    
    \item \textbf{Gradient aggregation error:} According to \eqref{eq6}, OTA transmitters scale their gradients with transmit power inversely proportional to the channel gain to ensure that the superimposed signals arrive with the desired weights. If a link suffers from deep fading, all the other devices sharing that channel must reduce transmit power, which amplifies the overall aggregation error and decelerates convergence \cite{XSB21}.
    
    \item \textbf{Network connectivity bottleneck:} The consensus speed of the approach in \eqref{Eq3} critically depends on the edge connectivity reflected in mixing weights $\Wv=[w_{nm}]$. Real-world D2D topologies often follow a “small-world’’ pattern: dense clusters connected by only a handful of inter-cluster edges. Consequently, some poorly connected nodes at the cluster edge become bottlenecks that dominate the global consensus rate.
\end{itemize}

\section{DFL via Device Clustering and Link Enhancement}

We develop a cluster-centric DFL model aggregation scheme to tackle the above limitations.
The core idea is to provision a small number of \emph{extra reliable communication links} between carefully selected devices (e.g., by backhaul connections), thereby strengthening the weakest parts of the network and accelerating global consensus. Because the placement of these links depends only on slow-varying large-scale channel statistics (e.g., large-scale fading coefficients), it can be optimized \emph{a priori} and kept fixed throughout training.

As shown in Fig.~\ref{fig: Overall framework}, the original D2D graph is viewed as a collection of densely connected \emph{intra-cluster} subgraphs bridged by sparsely connected \emph{inter-cluster} edges. We exploit this ``small-world" structure in two steps:
\begin{figure}[!t]
  \centering
\includegraphics[width=3.2 in]{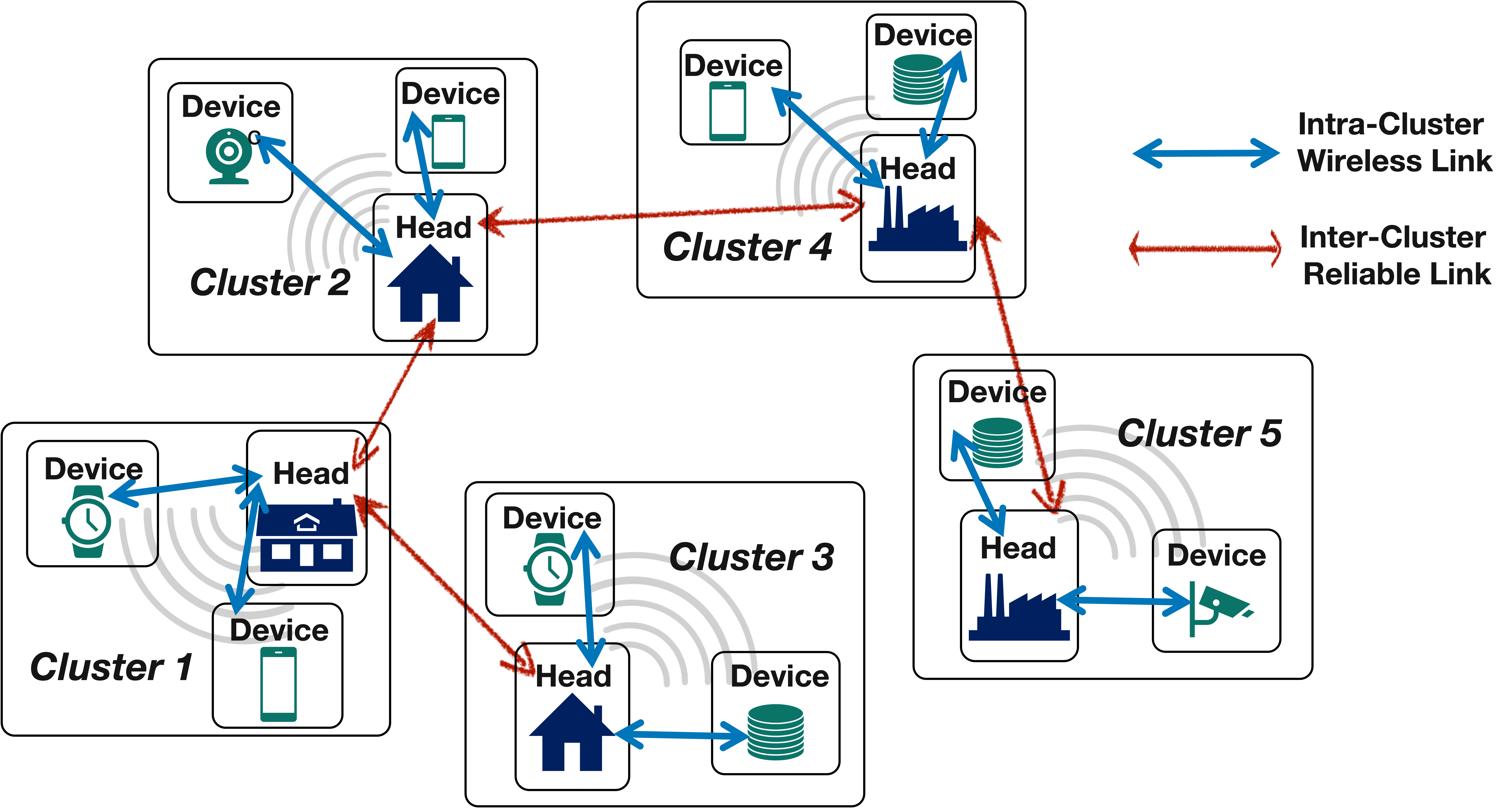}

  \caption{Illustration on the proposed DFL approach.}
  \label{fig: Overall framework}
\end{figure}

\begin{enumerate}[leftmargin=0.3cm]
    \item \textbf{Device clustering}: Using large-scale fading information, we partition the devices into clusters and pinpoint the few nodes whose poor connectivity in the intersections of clusters creates network bottlenecks.
    \item \textbf{Link enhancement and hybrid aggregation}: A limited budget of high-quality links is allocated to these stragglers, yielding robust shortcuts for inter-cluster gossip. Model aggregation is then decomposed into (a) noisy OTA exchanges \emph{within} each cluster and (b) reliable gossip-based exchanges \emph{across} clusters.
\end{enumerate}

This hybrid design simultaneously (a) alleviates the connectivity bottleneck by fortifying inter-cluster paths and (b) simplifies multiple-access scheduling, as gradient transmission now occurs alternately in intra-cluster and inter-cluster phases. 
\subsection{Device Clustering}

We represent the $N$-device D2D network by an undirected weighted graph $\mathcal{G}$ whose edge weights correspond to the large-scale fading coefficients. The goal is to partition $\mathcal{G}$ into $K$ disjoint clusters, where both $K$ and the cluster assignments are design variables. Denote by $\mathcal{C}_k$ the node set of cluster $k$ and by $v_k\in\mathcal{C}_k$ its \emph{cluster head} (centroid).
We jointly compute $K$, and $\{\mathcal{C}_k,v_k\}_{k=1}^{K}$ subject to the following requirements:
\begin{itemize}[leftmargin=0.3cm]
    \item \textbf{Star-type intra-cluster topology:} For every $k$, the subgraph induced by $\mathcal{C}_k$ must contain a star with $v_k$ at its center; that is, each node in $\mathcal{C}_k$ already shares a direct wireless link with $v_k$. This ensures that intra-cluster aggregation can adopt a server-based FL protocol with $v_k$ acting as the local ``server."
    
    \item \textbf{Link-enhancement budget:} We augment the connectivity among the heads $\{v_k\}$ by provisioning a limited number of reliable links (e.g., wired backhaul). Let $\mathcal{H}$ be the $K$-node weighted graph whose edges correspond to these links. Suppose that deploying a reliable transceiver at a node costs $L_{N}$, and creating a link of length $e(v_i,v_j)$ between $v_i$ and $v_j$ costs $L_{D}\cdot e(v_i,v_j)$. The total expense must satisfy
    \begin{align}\label{eq8}
        K L_{N}+L_{D}\sum_{1\le i<j\le K}e(v_i,v_j)\le B,
    \end{align}
    where $B$ is the overall budget and $e(v_i,v_j)$ is the edge weight in $\mathcal{H}$.  Moreover, each reliable link must obey the physical reach limit
    \begin{align}\label{eq9}
        \max_{1\le i<j\le K}e(v_i,v_j)\le \tau,
    \end{align}
    where $\tau$ is the maximum feasible distance.
    
    \item \textbf{Learning-rate maximization:} Subject to the topology and budget constraints above, we select $K$, $\{\mathcal{C}_k,v_k\}$, and $\mathcal{H}$ so that, under the hybrid aggregation protocol described next, the resulting DFL procedure achieves the largest possible learning rate.
\end{itemize}
Before formally formulating and solving the clustering problem, we first outline the OTA-based hybrid aggregation scheme over $\mathcal{G}$ and $\mathcal{H}$ in the sequel.

\subsection{Model Aggregation in the Clustered Network}

Given the clustering solution $\{\mathcal{C}_k,v_k\}$ and its head–link graph $\mathcal{H}$, the objective \eqref{eq: grand objective} is equivalent to 
\begin{equation}
    \min_{\xv\in\mathbb{R}^d}\;
    \sum_{k=1}^{K} p_k\,f_{\mathcal{C}_k}(\xv),
    \label{eq: grand objective2}
\end{equation}
where $p_k=D_{\mathcal{C}_k}/D$ with $D_{\mathcal{C}}=\sum_{i\in\mathcal{C}}|\mathcal{D}_i|$ and
\begin{equation}
    f_{\mathcal{C}_k}(\xv)=\sum_{i\in\mathcal{C}_k}
    \frac{|\mathcal{D}_i|}{D_{\mathcal{C}_k}}\,f_i(\xv).
    \label{eq: grand objective3}
\end{equation}
During any training iteration $t$ (index omitted for brevity), the model aggregation proceeds in two steps.  

\medskip
\noindent\textbf{Intra-cluster aggregation.}
Every device $i\in\mathcal{C}_k$ computes the stochastic gradient ${\nabla} f_{i}(\xv_k;\mathcal{B}_{i})$ and uploads it to the head $v_k$ via the OTA scheme in \eqref{eq5}. Using \eqref{eq_otareceiver}, the head recovers a noisy estimate of the weighted sum  $\sum_{i\in\mathcal{C}_k}\frac{|\mathcal{D}_i|}{D_{\mathcal{C}_k}}\,
    {\nabla} f_{i}(\xv_k;\mathcal{B}_{i})$, abbreviated as $\widehat{\gv}_k$, with the mixing weight $w_{iv_k}=\frac{|\mathcal{D}_i|}{D_{\mathcal{C}_k}}$. Then, it updates its intra-cluster model $\xv_k$ by using the estimated aggregated gradient, and broadcasts it to all members of $\mathcal{C}_k$.
    
\medskip
\noindent\textbf{Inter-cluster aggregation.}
For every $H>0$ training iterations, the heads perform a noiseless gossip step over the reliable links in $\mathcal{H}$ once:
\begin{align}
    \xv_k \;\gets\;
    \sum_{l\in\mathcal{N}_{\mathcal{H}}(v_k)}
    p_l\,\xv_l + (1-\sum_{l\in \mathcal{N}_\mathcal{H}(v_k)} p_l) \xv_k,
    \label{eq12}
\end{align}
where $\mathcal{N}_{\mathcal{H}}(v_k)$ denotes the neighbors of $v_k$ in $\mathcal{H}$. Restricting gossip to the $K$ heads, rather than to all the $N$ devices as in conventional DFL \cite{chocoSGD,XSB21}, markedly accelerates global consensus. The inter-cluster communication interval $H$ is introduced to reduce communication delay and boost the overall efficiency. The impact of $H$ is investigated in Sections~\ref{Sec_analysis} and \ref{sec_simulation}.

\medskip
\noindent The complete procedure is summarized in Algorithm~\ref{alg: SFLwG}. The proposed approach speeds up convergence by combining server-based OTA aggregation within clusters and infrequent, noiseless gossip among a small set of well-connected heads, at the cost of a modest additional deployment budget for the reliable links.

\begin{algorithm}[!t]
\caption{The proposed DFL algorithm.}
\label{alg: SFLwG}
\begin{algorithmic}[1]
\REQUIRE $\{\mathcal{C}_k,v_k,p_k\}_{k=1}^K$, $\{\eta_t\}_{\forall t}$, $H$, $\{h_{mn}^{(t)}\}_{\forall m,n,t}$, Initial value $\xv_{n}^{(0)} = \xv^{(0)},\forall n$.
\FOR{$t = 1$ to $T$}
    \FOR{$k \in [K]$  \textbf{in parallel}}
        \FOR{\textbf{each device} $i \in \mathcal{C}_k$ \textbf{in parallel}}
            \STATE  Compute ${\nabla} f_{i}(\xv_k^{(t)};\mathcal{B}_{i}^{(t)})$ via SGD.
            \STATE Upload the gradient to $v_k$ via \eqref{eq5}.
        \ENDFOR
    \ENDFOR
    \STATE $v_k$ estimates $\widehat{\gv}_k^{(t)}$ via \eqref{eq_otareceiver}.
    \STATE $v_k$ updates $\xv_k^{(t+\frac{1}{2})} \gets \xv_{k}^{(t)} - \eta_t \widehat{\gv}_k^{(t)}$.
    \IF{$t \mod H = 0 $}
        \STATE $\xv_k^{(t+1)} \gets \xv_k^{(t+\frac{1}{2})}+\sum_{l \in \mathcal{N}_{\mathcal{H}}(v_k)} p_l (\xv_l^{(t+\frac{1}{2})} - \xv_k^{(t+\frac{1}{2})} )  $.
    \ELSE
        \STATE $\xv_k^{(t+1)} \gets \xv_k^{(t+\frac{1}{2})}$.
    \ENDIF
    \STATE Each $v_k$ broadcasts $\xv_k^{(t+1)}$ to $\mathcal{C}_k$.
\ENDFOR
\STATE \textbf{Return} local models $\{\xv_n^{(T)}\}_{\forall n}$.
\end{algorithmic}
\end{algorithm}

\vspace{-0.4cm}
\section{Performance Analysis and System Optimization}
\label{sec: proposed method}

This section first establishes the convergence rate of the proposed clustered DFL algorithm and then uses the results to optimize the device clustering variables of $K,\mathcal{H}$, and $\{\mathcal{C}_k,v_k\}$.

\vspace{-0.6cm}
\subsection{Learning Convergence Analysis}
\label{Sec_analysis}
We begin with two standard assumptions, which are widely adopted in the distributed learning literature.

\noindent\textbf{Assumption 1.}  
Each local objective $f_n$ in \eqref{eq: local objective} is $L$-smooth and $\mu$-strongly convex, guaranteeing the existence and uniqueness of the global optimum $\xv^*$ of \eqref{eq: grand objective}.

\noindent\textbf{Assumption 2.}  
For every device $n$ and iteration $t$, the mini-batch gradient ${\nabla} f_{n}(\xv_n^{(t)};\mathcal{B}_{n}^{(t)})$ is unbiased and has bounded variance and second moment:
\begin{align*}
        &\Ebb \| {\nabla} f_{n}(\xv_n^{(t)};\mathcal{B}_{n}^{(t)}) - \nabla f_n (\xv_n^{(t)};\mathcal{D}_n) \|^2 \leq \sigma_n^2,\nonumber\\
&\Ebb \| {\nabla} f_{n}(\xv_n^{(t)};\mathcal{B}_{n}^{(t)}) \|^2 \leq G^2.
\end{align*}

Theorem \ref{thm: main} quantifies the MSE of the network-wide average model relative to $\xv^*$; its proof is omitted here. 

\enlargethispage{-\baselineskip} 
\begin{theorem}
\label{thm: main}
Suppose the instantaneous CSI $\{h_{mn}^{(t)}\}_{\forall m,n,t}$ is given at each iteration.
Let the SGD stepsize be $\eta_t=\frac{2}{\mu(t+a)}$ with a given constant $a=\Omega(H)>0$, and define $\overline{\xv}^{(t)}=\frac{1}{N}\sum_{n=1}^{N}\xv_n^{(t)}$ as the network-wide average model computed by Algorithm \ref{alg: SFLwG}.  It follows that\footnote{We corrected a typo in \eqref{theorem main} from the conference version.}
    \begin{align}    
          \Ebb \| \overline{\xv}^{(t)} &  - \xv^* \|^2 
       \leq  \frac{4 ( \sum_{k}p_k^2 \zeta_k^2 + \frac{1}{D^2}\sum_{n} |\mathcal{D}_n|^2 \sigma_n^2 ) t}{\mu^2 (t+a-2)^{2 } }\nonumber \\
          & + \frac{a^2 \| \xv^{(0)} - \xv^* \|^2}{(t+a-2)^{2 }} + \Ocal \left(H^2\frac{\log t}{t^2}  \right), \label{theorem main}
    \end{align}
where expectation is over mini-batch sampling and communication noise; and $\zeta_k^2 \triangleq \Ebb \left\| \widehat \gv_k^{(t)} - \sum_{i\in\mathcal{C}_k}\frac{|\mathcal{D}_i|}{D_{\mathcal{C}_k}}\,{\nabla} f_{i}(\xv_k;\mathcal{B}_{i}) \right \|^2$
is MSE caused by the intra-cluster OTA aggregation (cf. Line 8 of Algorithm \ref{alg: SFLwG}).
\end{theorem}

\subsection{Budget-Constrained Device Clustering Optimization}
\label{sec: partition scheme}

Theorem \ref{thm: main} shows that the dominant error terms decay as $\Ocal(1/t)$ and are shaped by the OTA aggregation term $\sum_{k}p_k^2 \zeta_k^{2}$.  Under the transceiver design in \eqref{eq6}, this error term depends on the instantaneous CSI $\{h_{mn}\}$ and, by \cite[Lemma 2]{HL21}, can be expressed as
\begin{align}
  \sum_{k=1}^K p_k^2  \zeta_k^2&\propto   \sum_{k=1}^K \left(\min_{i\in\mathcal{C}_k} \frac{|h_{iv_k}^{(t)}|^2}{|\mathcal{D}_i|^2}\right)^{-1}.
\end{align}
Since $h_{iv_k}^{(t)}\!\sim\!\mathcal{CN}(0,\alpha_{iv_k})$, $\min_{i}|h_{iv_k}^{(t)}|^{2}/|\mathcal{D}_i|^{2}$ is exponentially distributed with rate
$\sum_{i\in\mathcal{C}_k}|\mathcal{D}_i|^{2}/\alpha_{iv_k}$.
Motivated by this, without access to instantaneous CSI, we design the clusters by minimizing the following aggregation MSE statistic subject to the budget constraints in \eqref{eq8}–\eqref{eq9}:
\begin{align}
    \sum_{k=1}^K \left(\Ebb\left[\min_{i\in\mathcal{C}_k} \frac{|h_{iv_k}^{(t)}|^2}{|\mathcal{D}_i|^2}\right]\right)^{-1}=\sum_{k=1}^K\sum_{i\in \mathcal{C}_k}\frac{|\mathcal{D}_i|^2}{\alpha_{iv_k}}.\label{cut left tails}
\end{align}

We present a low-complexity, sub-optimal solution to the minimization of \eqref{cut left tails} as follows:
\begin{itemize}[leftmargin=0.3cm]
    \item \textbf{Construct $\mathcal{H}$ for a given $K$ and clustering.}
          With fixed $K$ and $\{\mathcal{C}_k,v_k\}$, the objective \eqref{cut left tails} is independent of $\mathcal{H}$.  
          We therefore build $\mathcal{H}$ at minimum cost by computing the minimum spanning tree (MST) of the $K$ heads, which takes $\Ocal(K^{2}\log K)$ time \cite{kleinberg2006algorithm}.

    \item \textbf{Cluster formation for a given $K$.}
          For any feasible $K\!\ge\!1$, the constrained clustering of $\{\mathcal{C}_k,v_k\}$ is NP-hard. We adopt a heuristic method to solve for a sub-optimal solution.  
          When the head vector $\boldsymbol{v}=[v_1,\dots,v_K]$ is fixed,  each node $i$ can be greedily assigned to the cluster that minimizes $|\mathcal{D}_i|^{2}/\alpha_{iv_k}$.  
          We then refine $\boldsymbol{v}$ iteratively via Gibbs sampling. Denote the head selection state in iteration $r$ as 
         $\boldsymbol{v}_r$. Consider all feasible neighboring states that differ from $\boldsymbol{v}_r$ in exactly one head index; we draw a new state $\boldsymbol{v}_{r+1}$ from the neighboring state with probability proportional to
          $e^{-{\rm Obj}(\boldsymbol{v}_r')/\Lambda_r}$, where ${\rm Obj}(\cdot)$ is the value of \eqref{cut left tails} evaluated at the neighboring state $\boldsymbol{v}_r'$, and $\Lambda_r=\rho\Lambda_{r-1}$ ($0<\rho<1$) is a decreasing ``temperature" parameter. Here, we adopt a cooling schedule for $\Lambda_r$, which balances the trade-off between exploration and exploitation. Finally, the best state encountered in $R$ iterations is returned.

    \item \textbf{Line search over $K$.}
          Starting from $K=1$, we iteratively increment $K$, apply the above two steps, and stop once the budget constraints are violated.  Finally, the best feasible solution is selected.
\end{itemize}
The complete procedure is summarized in Algorithm \ref{alg: lloyd}.

\begin{algorithm}[!t]
\caption{Device clustering and link construction.}
\label{alg: lloyd}
\begin{algorithmic}[1]
\REQUIRE $\{\alpha_{mn}\}_{\forall m,n}$, $\{|\mathcal{D}_n|\}_{\forall n}$, $R$, and $\{\Lambda_r\}$. 
\STATE Initialize $K=1$.
\WHILE{the constraints in \eqref{eq8}--\eqref{eq9} are not violated}
    \STATE Initialize a feasible $\vv_0$. Then, greedily associate nodes $\{\mathcal{C}_k\}$ to $\vv_0$ and compute an MST $\mathcal{H}$.
    \FOR{sampling iteration $r = 1,\cdots,R$}
        \FOR{each feasible neighboring state $\vv_r^\prime$ of $\vv_r$}
            \STATE Greedily associate nodes $\{\mathcal{C}_k\}$ to $\vv_r$ and compute an MST $\mathcal{H}$ w.r.t. $\vv_r$.
            \STATE Evaluate \eqref{cut left tails} w.r.t. $\vv_r^\prime$ as ${\rm  Obj}(\vv_r^\prime)$.
        \ENDFOR
        \STATE Draw $\vv_{r+1}$ from the neighboring states $\{\vv_r\}\cup \{\vv_r^\prime\}$ with a probability $\propto e^{-\frac{{\rm Obj}(\vv_r^\prime)}{\Lambda_r}}$.
    \ENDFOR
    \STATE Pick the best $\vv$ from $\{\vv_r\}_{r=1}^R$. $K\leftarrow K+1$
\ENDWHILE
\STATE \textbf{Return} the best solution of $K,\{\mathcal{C}_k\},\vv,$ and $\mathcal{H}$ with the minimum objective.
\end{algorithmic}
\end{algorithm}

\section{Numerical Results}
\label{sec_simulation}
\begin{figure*}[!t]
\centering
\begin{adjustbox}{scale=0.89,clip,trim=0cm 0cm 0cm 0cm}
\includegraphics[width=\linewidth]{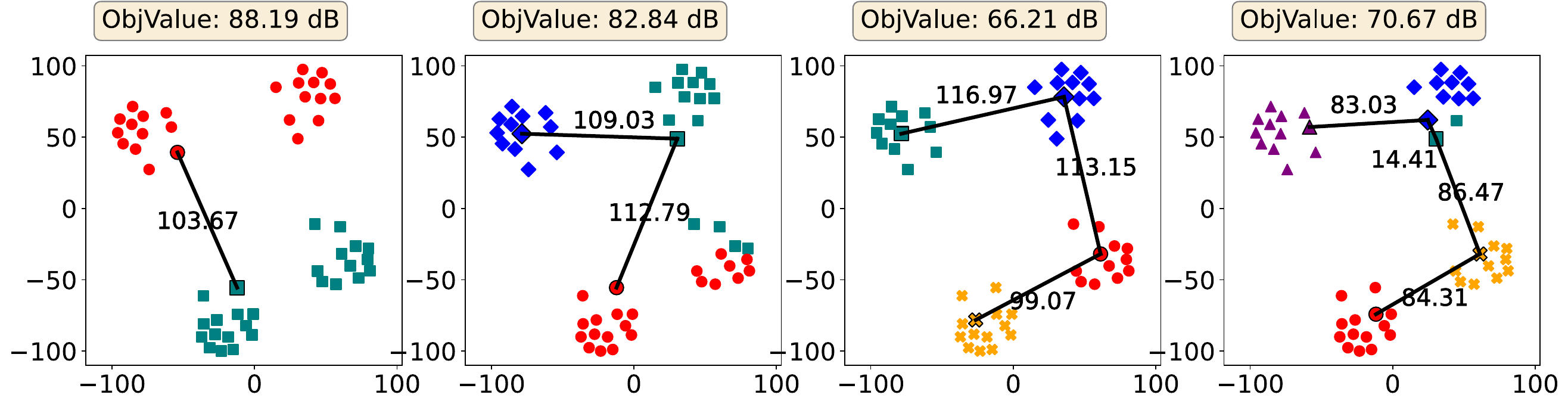}
\end{adjustbox}
\caption{Device clustering results and the objective (in dB) in Algorithm \ref{alg: lloyd} versus the value of $K \in \{2,3, 4, 5\}$. Different colors represent clustering groups $\{\mathcal{C}_k\}$. The reliable links in $\mathcal{H}$ is visualized via solid black lines with the link distances marked above.}
\label{fig: K2345}
\end{figure*}

\begin{figure}[t]
  \centering
  \scalebox{0.34}{\input{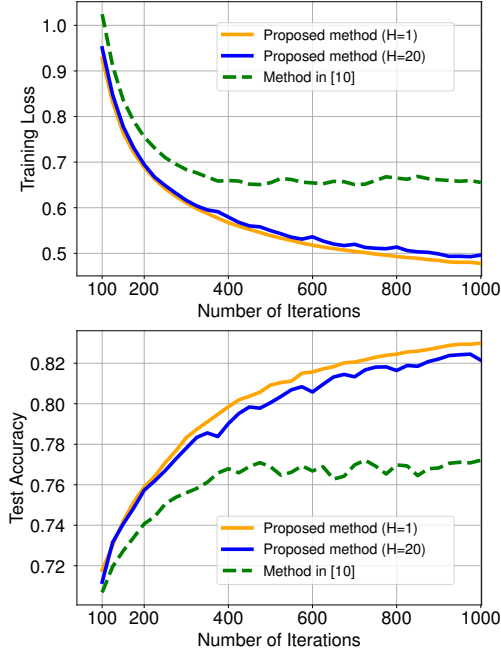}}
  \caption{Average training loss (top) and test accuracy (bottom) versus number of iterations for the Fashion-MNIST image classification with the D2D network in Fig. \ref{fig: K2345}.}
 \label{fig: compare}
\end{figure}

In this section we examine the proposed approach through simulations. A D2D network with $N=50$ devices is generated over a square region $[-100~\text{m},100~\text{m}]^{2}$ using a stochastic-block model with four blocks, where the intra- and inter-block connection probabilities are set to $0.99$ and $0.01$. The large-scale fading is set to $\alpha_{mn}=\alpha_0 d^{-\gamma}$ with the reference path loss $\alpha_0=0$ dB and path-loss exponent $\gamma=3.76$. The small-scale fading in $h_{mn}^{(t)}$ is drawn i.i.d. in each training iteration according to \eqref{eq_channel}. The transmit-power limit is $P_0=2$ W, and the communication noise power is $0$ dBW.

\smallskip
\noindent\textbf{Impact of the number of clusters.}  
 We first investigate the device clustering performance of Algorithm \ref{alg: lloyd} with budget parameters $\tau=120$, $B=530$, $L_N=50$, and $L_D=1$ in \eqref{eq8}–\eqref{eq9}. Fig.~\ref{fig: K2345} reports the clustering resulting and the objective value in \eqref{cut left tails} representing the convergence-rate error term versus $K$.  Among all feasible solutions, $K=4$ minimizes the objective.  The plot also shows that the objective is \emph{not} monotone in $K$: if $K$ grows too large, the budget constraints force cluster heads to lie closer together, which subsequently degrades the clustering performance.  Note that the line search in Algorithm~\ref{alg: lloyd} correctly identifies $K=4$ as optimal.

\smallskip
\noindent\textbf{Learning performance.}  
We next test the proposed clustered DFL method in Algorithm~\ref{alg: SFLwG} with $K=4$ on the Fashion-MNIST classification task \cite{Xiao2017}.  The $60{,}000$ training images are i.i.d. distributed among the $N=50$ devices, while the $10{,}000$ test images are used for evaluating the accuracy of the network-wide average model.  We train a $3$-layer multi-layer perceptron (MLP) model of $d=235,146$ parameters with stepsize $\eta_t=\tfrac{30}{1000+t}$, and local batch size $100$. We compare the proposed method with the baseline method in \cite{XSB21}, which relies on the OTA gossip scheme in \eqref{Eq3} for model aggregation. Fig.~\ref{fig: K2345} depicts the training loss and test accuracy.  Thanks to the reliable links between cluster heads, the proposed scheme mitigates inter-cluster stragglers and achieves markedly faster convergence than the baseline.  Moreover, the proposed method is insensitive to the value of the inter-cluster communication interval $H$. This observation corroborates the analysis in Theorem~\ref{thm: main}, where $H$ does not appear in the leading $\mathcal{O}(1/t)$ error term of the optimality gap.  From a practical standpoint, a larger $H$ is therefore desirable to reduce communication overhead without sacrificing accuracy.

\section{Conclusions}
\label{sec: conclusions}
In this work, we addressed the straggler problem that limits DFL model aggregation efficiency in large-scale D2D networks.  We proposed a budget-aware, cluster-centric framework, which identifies sensitive straggling devices and provisions a small set of reliable backhaul links between carefully chosen devices.  This design converts decentralized model aggregation into (i) fast device-to-server model aggregation within each cluster, and (ii) low-frequency gossip exchange among cluster heads.  We established an $\Ocal(1/t)$ convergence bound on the optimality gap, which explicitly captures the impact of model aggregation error. We then developed a budget-constrained device-clustering algorithm that minimizes the leading error term without requiring instantaneous CSI. Experiments on image classification verified that the proposed method markedly outperforms the state-of-the-art DFL approach in learning convergence. These results demonstrate that modest, strategically placed link enhancements can unlock substantial efficiency gains for large-scale DFL.

\bibliographystyle{IEEEtran}
\bibliography{IEEEabrv, journal}

\end{document}

%% file: preamble.tex
\usepackage[utf8]{inputenc}

\usepackage{amsmath}
\usepackage{amsfonts}
\usepackage{amssymb}
\usepackage{textcomp}
\usepackage{color}
\usepackage{xcolor}
\usepackage{amsthm}
\usepackage{bm}
\usepackage{mathrsfs}
\usepackage{graphicx}
\usepackage{subcaption}
\usepackage{epsf}
\usepackage{xpatch}
\usepackage{setspace} 
\usepackage{enumitem,kantlipsum}
\usepackage{multirow,booktabs}
\usepackage{latexsym}
\usepackage{accents}
\usepackage[export]{adjustbox}
\usepackage{dsfont}
\usepackage{mathcomSTEv4}
\usepackage{url}

\usepackage{tikz}
\usetikzlibrary{arrows,shapes,chains,matrix,positioning,scopes,patterns,calc,
decorations.markings,
decorations.pathmorphing,
}

\usepackage{pgfplots}
\pgfplotsset{compat=1.3}
\usepgflibrary{shapes}

\theoremstyle{definition}

\usepackage{amsfonts}
\usepackage{fancyhdr}
\usepackage{comment}
\usepackage{amsmath}
\usepackage{changepage}
\usepackage{stfloats}
\usepackage{amssymb}
\usepackage{graphicx}
\usepackage{color}
\usepackage{xcolor}
\usepackage{indentfirst}
\usepackage[font={small}]{caption}
\usepackage{tkz-graph}
\usetikzlibrary{graphs,graphs.standard}

\newtheorem{theorem}{Theorem}

\usepackage{mathtools}